\documentclass[sigconf,screen]{acmart}
\usepackage{multirow}
\usepackage[ruled,vlined,lined,commentsnumbered]{algorithm2e}
\usepackage{colortbl}
\usepackage{enumitem}
\usepackage{makecell}
\usepackage{subfig}
\usepackage{tcolorbox}
\tcbuselibrary{breakable}

\usepackage[normalem]{ulem}
\usepackage{microtype}

\AtBeginDocument{%
  }

\setcopyright{cc}
\setcctype{by}
\acmDOI{10.1145/3843282.3844441}
\acmYear{2026}
\copyrightyear{2026}
\acmISBN{979-8-4007-2985-0/2026/10}
\acmConference[AgenticDev '26]{Proceedings of the 1st International Workshop on Agentic AI for Next-Generation Software Development}{October 12--16, 2026}{Munich, Germany}
\acmBooktitle{Proceedings of the 1st International Workshop on Agentic AI for Next-Generation Software Development (AgenticDev '26), October 12--16, 2026, Munich, Germany}
\acmSubmissionID{asews26agenticdevmain-p126-p}
\received{2026-07-24}
\received[accepted]{2026-08-20}

\begin{document}

\title[Teach-to-Crash]{Teach-to-Crash: A Closed-Loop Student–Teacher LLM Framework for Collision-Inducing Test Scenario Generation}

\author{Zaid Ghazal}
\correspondingauthor
\orcid{0009-0000-5960-1765}
\affiliation{%
  \institution{University of Michigan-Dearborn}
  \city{Dearborn}
  \state{MI}
  \country{USA}
}
\email{zghazal@umich.edu}

\author{Khouloud Gaaloul}
\orcid{0000-0002-4156-9768}
\affiliation{%
  \institution{University of Michigan-Dearborn}
  \city{Dearborn}
  \state{MI}
  \country{USA}
}
\email{kgaaloul@umich.edu}

\author{Bruce Maxim}
\orcid{0000-0002-0979-7787}
\affiliation{%
  \institution{University of Michigan-Dearborn}
  \city{Dearborn}
  \country{USA}
}
\email{bmaxim@umich.edu}

\renewcommand{\shortauthors}{Ghazal, Gaaloul, Maxim}

\begin{abstract}
Validating Autonomous Driving Systems (ADS) in simulation requires testing architectures that can discover rare, safety-critical failures while generating scenarios that are executable, diverse, and useful for downstream failure analysis. We introduce \textit{Teach-to-Crash}, a closed-loop testing framework that combines a constrained ego-centric scenario representation, stagnation-aware search control, and a dual-LLM architecture for adaptive failure discovery. A high-reasoning Teacher LLM acts as an adaptive search controller, while a low-reasoning Student LLM emits simulator-executable scenarios in a strict JSON schema. The Teacher intervenes only when rolling collision rate and time-to-collision metrics stagnate, providing strategic guidance to redirect the search. In a CARLA case study with two experimental setups that vary the ego vehicle's speed policy, \textit{Teach-to-Crash} achieves the highest Collision Hit Rate ($90.79\%$), the shortest mean Time-to-Collision ($18.31s$), and a competitive Collision Discovery Rate ($136.21$). PAFOT attains a higher mean CDR ($179.44$), but with substantially larger variance. \textit{Teach-to-Crash} also yields the highest diversity ($0.547$) and, averaged across both setups on the CARLA Traffic Manager controller, the highest avoidability-based usefulness proxy ($60.04\%$) among the compared methods. These results, within the evaluated CARLA scope, provide evidence that closed-loop dual-LLM reasoning can steer adversarial simulation-based testing over a constrained executable program space, generating failures that are frequent, structurally diverse, and assessed as more frequently avoidable.
\end{abstract}



\maketitle

\section{Introduction}
\label{sec:intro}

Simulation-based testing is essential for evaluating autonomous driving systems (ADS); it enables controlled variation of the operating environment, repeatable measurement, and replayable analysis of rare hazardous interactions in high-fidelity simulators and scenario languages~\cite{li2020av,cheng2023behavexplor,gambi2019automatically,fremont2019scenic}. The core difficulty is that safety-critical multi-agent behaviors occupy a high-dimensional space where failures are rare and small changes in non-ego behavior can flip an interaction from benign to collision-inducing~\cite{abdessalem2018testing,cheng2023behavexplor,li2020av}.

Search-based methods optimize scenario parameters via evolutionary search, fuzzing, and surrogate assistance~\cite{wegener2004evaluation,buehler2005evolutionary,abdessalem2018testing,deb2002fast,calo2020generating,li2020av,cheng2023behavexplor,gambi2019automatically,fremont2019scenic,jin2011surrogate,haq2022efficient,dreossi2019compositional,kolb2021fitness}, while learning-based methods train adversarial or generative agents over risky interactions~\cite{abeysirigoonawardena2019generating,koren2018adaptive,feng2021intelligent,chen2021adversarial,wachi2019failure,kuutti2020training,rempe2022generating,zhong2022guided}. Both approaches improve search effectiveness but rely on low-level parameterizations requiring domain expertise. Consequently, recent research has begun leveraging LLMs to generate, refine, and diversify scenarios from natural-language intent to enable scalable scenario construction. LLM-based methods generate, rewrite, or diversify scenarios from higher-level intent~\cite{ChatScene,li2024chatsumo,AgentsLLM,OmniTester,LLMAttacker,SeekingToCollide,gaaloul2026grammar}.

Across these families, two bottlenecks remain persistent under fixed budgets. First, scenario representation can waste executions on behaviors that never become ego-relevant, and stagnation can trap generation in near-duplicate interactions after it reaches a promising region~\cite{cheng2023behavexplor,li2020av,ji2025autonomous,pafotpaper}. Second, existing LLM-based methods primarily rely on a single open-loop prompt-engineering framework for scenario generation, and mitigate well-known issues such as hallucinations and inconsistent constraint satisfaction through post-hoc validation or repeated regeneration when scenarios are ineffective. Open-loop LLM-based test generation can translate high-level intent into executable programs, yet without execution feedback it cannot determine whether the resulting behavior actually improves the collision rate or merely changes the text of the scenario. In contrast, multi-agent LLM architectures for ADS testing remain underexplored, particularly when specialized agents collaborate to guide adaptive search over an executable scenario space instead of relying on a single prompt. This makes closed-loop multi agent generation, where LLMs exchange feedback and adapt to scenario outcomes, a promising frontier for discovering critical ADS test scenarios.

To address these bottlenecks, we propose \textit{Teach-to-Crash}, a closed-loop Student--Teacher LLM architecture that generates failure inducing ADS test scenarios through prompt engineering over a configurable, ego-centric, position-based encoding. Under a fixed testing budget, \textit{Teach-to-Crash} aims to discover simulator-executable scenarios that maximize the collision rate, shorten the time-to-collision, avoid collapsing onto near-duplicate failures, and remain useful for downstream failure analysis.

The Student was selected as a lower-capability tier by design, reserving the higher-reasoning Teacher for infrequent strategic interventions rather than every generation step. Without execution feedback, such models tend to produce repetitive high-probability scenarios, limiting exploration of more diverse and challenging cases. This motivates the Teacher's role in guiding subsequent refinement \cite{evidence_self_refinement, opro}. Accordingly, a second, higher-reasoning-tier Teacher LLM closes the loop, activating only when progress in collision rate and time-to-collision stalls to issue stagnation-aware guidance, informed by execution feedback, that redirects the Student toward higher-collision-rate scenarios; this split lets the lightweight model generate scenarios in bulk while reserving stronger reasoning for the rare strategic-guidance calls. Our evaluation does not isolate this role separation from model capability. We evaluate only the proposed closed-loop framework against the Student-only baseline (Section~\ref{subsec:rq1}). We do not compare against a same-model closed-loop configuration or a single, more powerful LLM executing the full workflow, and therefore make no claims regarding these alternatives. Such comparisons are left for future work (Section~\ref{sec:discussion}). Because LLM outputs may violate syntactic or semantic constraints, a scenario validator checks both LLMs' outputs structurally, discarding invalid responses and clipping admissible values before execution.


We evaluate \textit{Teach-to-Crash} in a CARLA case study on Town06, a low-density, highway-oriented map with four-to-six-lane road configurations, across two experimental setups, against ChatScene and PAFOT. Averaged across both setups, \textit{Teach-to-Crash} achieves the highest mean Collision Hit Rate ($90.79\%$ vs.\ $79.09\%$ for PAFOT and $64.03\%$ for ChatScene), the shortest mean Time-to-Collision ($18.31$ s vs.\ $23.71$ s and $30.32$ s), the most diverse failure archive ($404.4$ unique clusters per run, unique-failure ratio $0.900$), and the highest avoidable collision rate under our usefulness proxy ($60.04\%$, vs.\ $39.80\%$ on Setup~A alone with the CARLA Traffic Manager and $80.27\%$ on Setup~B). Its Collision Discovery Rate averages $136.21$ collisions per simulated hour, over $3\times$ ChatScene's, remaining competitive with PAFOT ($179.44$ mean CDR but substantially larger variance). These results indicate that, within this evaluated scope, dual-LLM reasoning is associated with higher failure coverage, diversity, and avoidability-based usefulness.

\noindent\textbf{Contributions.} The paper makes four contributions.
\begin{enumerate}
    \item We propose a closed-loop, dual-LLM Student--Teacher architecture for collision-inducing ADS test scenario generation, evaluated within the demonstrated CARLA scope (Section~\ref{sec:eval}). Unlike the single-LLM, open-loop generators we compare against (Section~\ref{sec:related}), it uses a second, higher-reasoning-tier LLM as a dedicated search controller invoked only on stagnation.
    \item We propose an adaptive Teacher LLM that responds to detected plateaus in failure discovery by revising the Student's generation policy. Our ablation associates this mechanism with the reported share of discovered collisions, without isolating its causal contribution or intervention timing (Section~\ref{sec:discussion}).
    \item We formulate an ego-centric, position-based scenario program representation that constrains the search space for reliable LLM generation while preserving expressive multi-step interactions. We also introduce a scenario validator that enforces structural constraints on every LLM output before execution.

    \item We evaluate \textit{Teach-to-Crash} in CARLA against two representative baselines across two experimental setups and two ego controllers: CARLA's built-in autopilot/Traffic Manager and the TransFuser controller~\cite{Chitta2023PAMI}.
\end{enumerate}

\section{Motivating Example and Problem Definition}
\label{sec:problem}

\begin{figure}[t]
    \centering
    \includegraphics[width=1\linewidth]{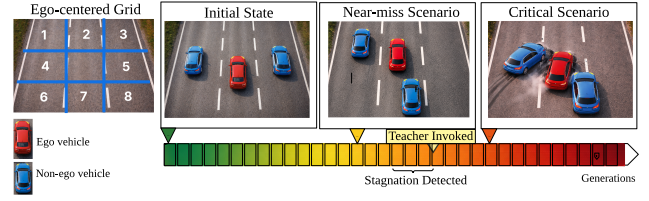}
    \caption{Running example of the ADS under test using CARLA simulator.}
    \Description{Left: a legend showing the 3x3 ego-centered position grid with cells numbered 1-8 surrounding the ego vehicle's central cell, plus a key marking the ego vehicle in red and non-ego vehicles in blue. Right: three top-down CARLA highway snapshots showing one red ego vehicle and two blue NPC vehicles progressing from an Initial State with the vehicles spaced apart, to a Near-miss Scenario with the vehicles converging, to a Critical Scenario showing a collision with skid marks. Below the snapshots, a green-to-red gradient timeline labeled Generations marks these three states in sequence and highlights a Stagnation Detected region where the Teacher is invoked, after which the search proceeds to the collision.}
    \label{fig:running}
\end{figure}

To motivate our approach, consider a squeeze-and-cut-in maneuver tested by a single LLM in an open-loop CARLA~\cite{Dosovitskiy2017CARLA} setting: two adversarial vehicles face an ego vehicle, and the LLM places one front-left and one rear-right of it, gradually sharpening the interaction into a near-miss where the ego brakes to avoid the cut-in. The search then stagnates at this local optimum, repeatedly generating similar near-collisions—a recognized challenge in ADS testing~\cite{cheng2023behavexplor,ji2025autonomous}: without execution feedback, the generator cannot tell whether it is progressing, converging prematurely on a narrow interaction pattern~\cite{LLMAttacker}, which recent work addresses via LLM reasoning to escape local optima~\cite{LeGEND2024}.

Our goal is to generate collision-inducing scenarios that are safety-critical, structurally diverse, and useful for downstream debugging via avoidable failures, using a closed-loop Student--Teacher framework: stagnation is detected via rolling collision-rate and time-to-collision metrics, and the Teacher LLM responds with a strategic directive steering the search toward more collision-inducing, diverse scenarios. Because reaching a collision often takes substantial time under a fixed budget, fewer critical cases are found; we address this via an ego-centered position grid with a fixed number of positions per vehicle, a configurable space compilable into continuous scenarios, defined next with the resulting optimization problem (usefulness approximated via the avoidability-based proxy in Definition~\ref{def:avoidability}).

Let $\mathbb{T}=[0,H]$ be a simulation time horizon. An \emph{ego-centered position grid} is a finite set of discrete position identifiers $\mathcal{P}(N)=\{1,\ldots,N\}$ and a decoding function $\mathit{DecodePos}$ mapping each $p\in\mathcal{P}(N)$ to a continuous target region in the ego's local frame, whose center is the target waypoint for a non-ego vehicle; larger $N$ gives finer placement control. As in Figure~\ref{fig:running}, we use $N=9$, partitioning the ego neighborhood into nine regions: positions $1$--$8$ for NPCs, position $9$ reserved for the ego.

\noindent\textbf{Example.} Consider \textsf{npc0} starting at position~$1$ (front-left) with target speed $32$\,mph and shifting toward position~$2$ (ahead), while \textsf{npc1} starts at position~$5$ (right) at $28$\,mph and drifts toward position~$4$ (front-right); the encoding gives \textsf{npc0}: $[32,1]$, $[32,2]$ and \textsf{npc1}: $[28,5]$, $[28,4]$. If the metrics stagnate, the Teacher may instruct the Student to tighten \textsf{npc0}'s cut-in while positioning \textsf{npc1} as a blocker, yielding earlier and more frequent collisions.


\begin{definition}[\textbf{Position-Based Scenario}]
\label{def:scenario}
Fix an integer $K\geq 2$ denoting the number of positions distributed over $\mathbb{T}=[0,H]$ at a uniform interval $\Delta=\frac{H}{K-1}$. For a non-ego vehicle $i$, a \emph{position} at step $k\in\{0,\ldots,K-1\}$ is a pair $\ell(i,k)=\langle p(i,k), v(i,k)\rangle$, where $p(i,k)\in\mathcal{P}(N)$ is a discrete position identifier and $v(i,k)\in [v_{\min},v_{\max}]$ is a target speed.

Let $M$ be the number of non-ego vehicles. A \emph{position-based scenario} is a tuple $\theta=\big(\mathcal{L}(1),\ldots,\mathcal{L}(M)\big)$,
where each $\mathcal{L}(i)$ is a length-$K$ sequence of position pairs,
$\mathcal{L}(i)=\big(\ell(i,0),\ell(i,1),\ldots,\ell(i,K-1)\big)$.
We denote the set of all encoded scenarios by $\mathcal{S}(N,K)$.
\end{definition}

Given $\theta\in\mathcal{S}(N,K)$, its \emph{concretization} is the executable program that, for each vehicle $i$ and step $k$, decodes $p(i,k)$ via $\mathit{DecodePos}$ into a target waypoint and drives the vehicle toward it while tracking speed $v(i,k)$ over $[k\Delta,(k+1)\Delta]$.

\begin{definition}[\textbf{Scenario Criticality}]
\label{def:criticality}
Executing $\theta$ yields a verdict $y(\theta) \in \{\mathit{pass}, \mathit{fail}\}$, where $\mathit{fail}$ indicates a collision. We also define a \emph{criticality score} $\mathrm{Crit}(\theta)$ quantifying risk even absent a collision, adapting PAFOT's composite score~\cite{pafotpaper}, used there to guide generation toward more critical outcomes. $\mathrm{Crit}(\theta)$ aggregates four normalized, direction-corrected surrogate risk metrics via Eq.~\eqref{eq:crit}, scaled so larger values indicate greater criticality:
\begin{equation}
\label{eq:crit}
\mathrm{Crit}(\theta) =
w_{1} \cdot f_{\mathrm{METTC}}(\theta)
+ w_{2} \cdot f_{\mathrm{MD}}(\theta)
+ w_{3} \cdot f_{\mathrm{SD}}(\theta)
+ w_{4} \cdot f_{\mathrm{ET}}(\theta),
\end{equation}
$\mathrm{METTC}$ (s) is the minimum estimated time-to-collision over execution; $\mathrm{MD}$ (m) the minimum Euclidean distance between ego and NPC; $\mathrm{SD}$ (m) the safety distance a vehicle should maintain from traffic, i.e., how far the trajectory falls below this margin rather than raw closest-approach distance; and $\mathrm{ET}$ (s) elapsed time to collision within the rollout horizon—smaller raw values indicate greater criticality for all four, computed for every scenario regardless of outcome (non-colliding rollouts take $\mathrm{METTC}$/$\mathrm{MD}$ from closest approach, $\mathrm{ET}$ from the full horizon). The weights $w_{1}, \dots, w_{4}$ are positive and fixed in the evaluated configuration, larger $\mathrm{Crit}(\theta)$ indicating more safety-critical outcomes.
\end{definition}

As discussed in Section~\ref{sec:monitoring}, the online loop does not optimize $\mathrm{Crit}(\theta)$ directly; the Teacher's intervention is instead triggered by the cheaper rolling collision rate and TTC in Eq.~\eqref{eq:rolling_metrics}. The richer $\mathrm{Crit}(\theta)$ aggregates additional signals not tracked online, retained in the archive for offline analysis rather than as the live control signal.

\begin{definition}[\textbf{Scenario Diversity}]
\label{def:diversity}
Let $G=\{\theta_1,\ldots,\theta_n\}$ be a set of generated scenarios. For each scenario $\theta$ with $K$ positions, define the fingerprint
\[
F(\theta)=\bigl\{(i,\,k,\,p(i,k),\,b(v(i,k)))\;:\;i\in\{1,\ldots,M\},\;k\in\{0,\ldots,K{-}1\}\bigr\},
\]
where $b(v)=\lfloor v/10\rfloor$ maps a target speed in $[0,60]$ mph to one of seven speed bins, preventing minor speed perturbations from being treated as behaviorally distinct.

The \emph{diversity} of $G$ is the mean pairwise Jaccard distance between scenarios:
\[
\mathrm{Div}(G)=\frac{2}{|G|(|G|-1)}\sum_{1\le i<j\le|G|}1-\frac{|F(\theta_i)\cap F(\theta_j)|}{|F(\theta_i)\cup F(\theta_j)|}.
\]
Since $G$ has $|G|(|G|-1)/2$ unordered pairs, the prefactor $2/(|G|(|G|-1))$ is the reciprocal pair count, so $\mathrm{Div}(G)$ is simply the \emph{average} pairwise distance over $G$; larger values indicate the set covers more structurally distinct interaction patterns.
\end{definition}

\begin{definition}[\textbf{Avoidable Collision}]
\label{def:avoidability}
Let $\theta$ be a scenario ending in a collision at time $t_c$. We sample the ego's kinematic state at discrete instants in the pre-collision window $[t_c-2,\,t_c]$. The collision is \emph{avoidable} if, at any instant, either (i) at least $5$\,m of lateral space is available for the ego to maneuver into, or (ii) the clear distance ahead exceeds the required stopping distance $d_{\text{stop}} = v^2/(2a) + 1.5$, where $v$ is the ego's speed, $a=6.5$\,m/s$^2$ an assumed maximum deceleration, and $1.5$\,m a fixed safety margin. Otherwise, the collision is \emph{unavoidable}. The $2$\,s pre-collision window, $5$\,m lateral threshold, $6.5$\,m/s$^2$ deceleration, and $1.5$\,m margin are implementation-chosen heuristic parameters, not values drawn from an external validated standard or citation. This kinematic proxy is not a substitute for reachability analysis or controller-specific reasoning; it indicates only that a generic controller had room or time to avoid the collision. We treat the resulting \emph{avoidable collision rate} as an \emph{avoidability-based usefulness proxy}, weak evidence that a scenario exposes a correctable ADS weakness rather than a physically unavoidable impact.
\end{definition}

\noindent\textbf{Critical Scenario Generation Problem.}
Let ADS be the system under test, and let $S(N,K)$ denote the scenario space induced by an ego-centered position grid with $N$ discrete positions and $K$ positions per non-ego vehicle. Given a testing budget $B$ (e.g., simulation time or rollout count), scenario generation selects and executes $G \subseteq S(N,K)$ with total execution cost at most $B$ while (i) maximizing the fraction of failure-inducing scenarios, (ii) maximizing their criticality, and (iii) maximizing diversity. Formally, we approximately solve:

\[
\begin{aligned}
\max_{G \subseteq S}\Bigg(
&\frac{1}{|G|}\sum_{\theta \in G}
    \mathbb{I}[y(\theta)=\mathit{fail}], \\
&\frac{1}{|G|}\sum_{\theta \in G}
    \mathrm{Crit}(\theta),
\;\mathrm{Div}(G)
\Bigg)
\quad
\text{s.t.}\quad \mathrm{cost}(G)\le B.
\end{aligned}
\]

As noted after Definition~\ref{def:criticality}, the online loop tracks failure fraction and criticality via the cheaper rolling collision-rate/TTC statistics rather than $\mathrm{Crit}(\theta)$; diversity is likewise not tracked online, instead encouraged via prompts and evaluated offline over the full archive (Section~\ref{sec:monitoring}).

\section{Related Works}
\label{sec:related}

This section reviews search-based, learning-based, and LLM-driven work on critical scenario generation for ADS.

\textbf{Search-Based Critical Scenario Generation.} Search-based testing formulates critical scenario generation as optimization over scenario parameters. It steers execution toward failures via evolutionary search, fuzzing, and surrogate guidance~\cite{wegener2004evaluation,buehler2005evolutionary,ben2016testing,abdessalem2018testing,deb2002fast,calo2020generating,li2020av,cheng2023behavexplor,gambi2019automatically,fremont2019scenic,jin2011surrogate,haq2022efficient,dreossi2019compositional,kolb2021fitness}. PAFOT~\cite{pafotpaper} is closest to our work: it introduces the same 9-position, ego-centric grid and uses a single-objective genetic algorithm over METTC, minimum distance, safety distance, and execution time (Definition~\ref{def:criticality}) to lower estimated time-to-collision. \textit{Teach-to-Crash} replaces this genetic search with LLM-driven generation and stagnation-aware Teacher rewriting, and uses PAFOT as its primary non-LLM baseline (Section~\ref{subsec:baselines}). AVFuzzer~\cite{li2020av} similarly searches an ego-relative interaction space but targets TTC/minimum-distance directly through fuzzing rather than a discrete, LLM-driven grid. BehAVExplor~\cite{cheng2023behavexplor} and DoFuzz~\cite{ji2025autonomous} instead prioritize \emph{behavioral diversity} explicitly, clustering or scoring novelty online to steer away from visited regions. \textit{Teach-to-Crash} relies instead on prompt-level diversity instructions (Section~\ref{sec:monitoring}), so our evaluation lacks a diversity-aware search baseline (Section~\ref{sec:discussion}). Other directions accelerate search via reinforcement learning and adversarial generation~\cite{humeniuk2024reinforcement,goss2022eagle,zheng2020rapid}, but still depend on hand-designed operators.

\textbf{Deep Learning for Safety-Critical Scenario Generation.} Deep learning approaches train adversarial or generative agents that shape traffic toward failures~\cite{abeysirigoonawardena2019generating,koren2018adaptive,feng2021intelligent,chen2021adversarial,wachi2019failure,kuutti2020training}, and others learn priors over risky interactions for controllable sampling~\cite{rempe2022generating,zhong2022guided}. Feng et al.~\cite{feng2021intelligent} learn a dense, naturalistic-and-adversarial importance-sampling representation rather than a fixed discrete grid. This offers finer-grained adaptivity, but at the cost of a trained sampling model rather than a prompt-driven one. These approaches can be effective in large continuous spaces but typically require substantial simulation budgets.

\textbf{LLM-Driven Critical Scenario Generation.} LLM-driven work uses language models to synthesize, rewrite, or diversify scenarios from higher-level intent. ChatScene~\cite{ChatScene} retrieves and assembles Scenic code fragments from natural-language descriptions into executable scenarios, and serves as our leading open-loop LLM baseline (Section~\ref{subsec:baselines}). ChatSUMO~\cite{li2024chatsumo} similarly targets SUMO rather than a high-fidelity simulator. LeGEND~\cite{LeGEND2024} uses an LLM in a top-down, retrieval-and-refinement pipeline over existing scenario templates. Like our work, it identifies single-pass LLM generation as prone to local optima. Unlike \textit{Teach-to-Crash}, however, it does not use a second, higher-capacity LLM as a dedicated online controller invoked under a stagnation criterion. Other methods rewrite existing scenes or induce adversarial behaviors through generated strategies and trajectories~\cite{AgentsLLM,OmniTester,LLMAttacker,SeekingToCollide}. These approaches raise automation but use the LLM in a single-stage role with separate validation or regeneration, rather than closing the loop with execution feedback. TARGET~\cite{TARGET2025} instead uses an LLM to extract traffic-rule knowledge into a compositional domain-specific language, validating each component before synthesizing executable simulation scripts. Unlike \textit{Teach-to-Crash}, its representation is derived from traffic-rule text rather than an ego-centric position/speed grid, and its pipeline performs single-pass, validated generation rather than closed-loop, stagnation-aware search. \textit{Teach-to-Crash} instead uses LLM reasoning inside the testing loop as adaptive search control over a constrained executable scenario-program space. Our comparison set is representative rather than exhaustive. Across both families, evaluation tends to emphasize aggregate collision yield, whereas our study also measures diversity and usefulness.

\section{Methodology}
\label{sec:approach}


\noindent\textbf{Algorithm parameters and terminology.} A \emph{scenario} $\theta$ is the position-based program defined in Definition~\ref{def:scenario}; a \emph{rollout} is one simulated execution of that scenario in CARLA; a \emph{batch} is the $B$ scenarios produced by one \textsc{StudentGenerate} call; an \emph{iteration} is one pass of the \textbf{for} loop in Algorithm~\ref{alg:teach-to-crash}; and a \emph{run} is one complete execution of Algorithm~\ref{alg:teach-to-crash} from $i=1$ to termination, averaged over $10$ independent runs per method/setup (Section~\ref{sec:exp-setup}). We set $B=10$, $W=10$, $\mathrm{CR}_{\text{target}}=90\%$, and $\mathrm{TTC}_{\text{target}}=12$\,s (Section~\ref{subsec:baselines}), and set $I_{\max}=45$ for each run; NPC bounds are position IDs in $[1,8]$ and speeds in $[0,60]$\,mph (Section~\ref{sec:scenario_representation}). \textsc{BudgetExhausted} reduces to $i>I_{\max}$: a run always executes $I_{\max}$ iterations without early termination once targets are reached, since these targets only gate Teacher intervention through \textsc{Stagnant} (Eq.~\eqref{eq:stagnation}), so the abstract budget $B$ of Section~\ref{sec:problem} is realized as at most $I_{\max}\times B$ scenarios per run. Table~\ref{tab:rq1-teacher-ablation} reports the realized totals, lower since scenarios failing validation are skipped.

The \emph{outcome history} $\mathcal{H}$ is the growing, append-only sequence of $(s,o)$ pairs accumulated over a run, ordered by iteration index, each $o$ recording the collision verdict $y(\theta)$, the observed or proxy $\mathrm{TTC}$, the minimum ego--NPC distance, the criticality score $\mathrm{Crit}(\theta)$ (Definition~\ref{def:criticality}), the avoidability label (Definition~\ref{def:avoidability}) when applicable, the iteration index, the prompts and responses, the simulator logs, and the Definition~\ref{def:diversity} fingerprint. $\mathcal{H}$ is never truncated within a run; its most recent $W{=}10$ records, $\mathcal{S}_W$, feed Eq.~\eqref{eq:rolling_metrics} and are rendered into the prompts, bounding prompt length as $\mathcal{H}$ grows.

\subsection{Framework Overview}
\label{sec:framework_overview}

\begin{figure*}
    \centering
    \includegraphics[width=0.99\linewidth]{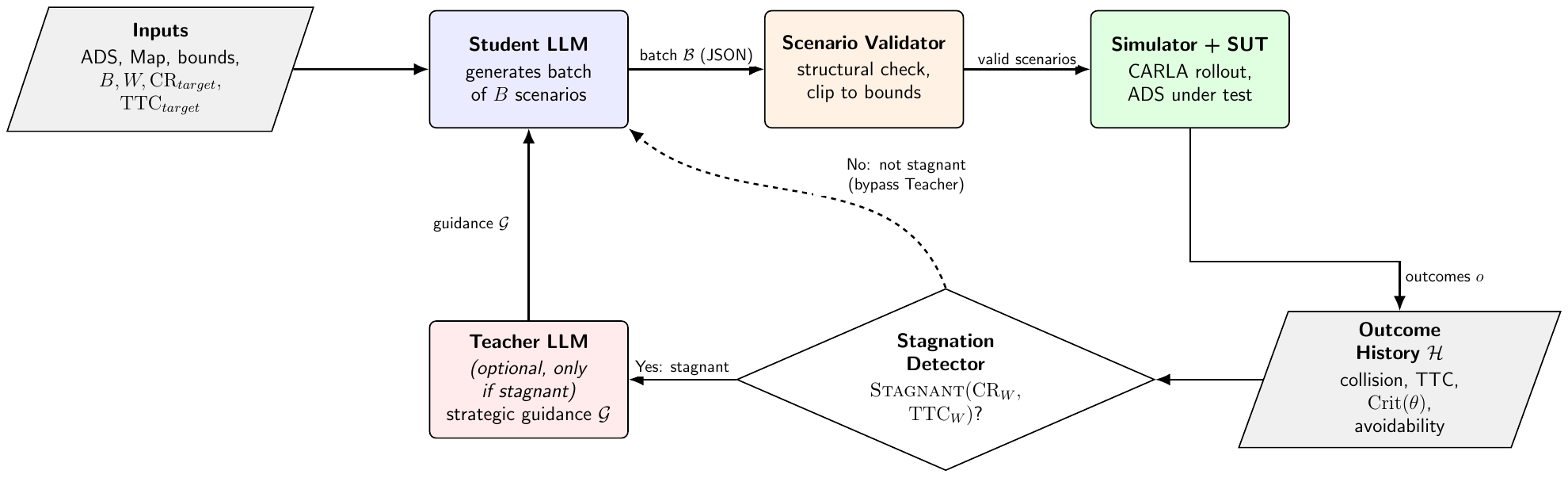}
    \caption{Teach-to-Crash workflow. Inputs (ADS, map, bounds, search parameters) start the loop at the Student LLM, whose JSON batch $\mathcal{B}$ passes through the Scenario Validator before execution. Outcomes, produced by the Simulator/SUT, are appended to the Outcome History $\mathcal{H}$. The Stagnation Detector consumes $\mathcal{H}$ each iteration: if \textsc{Stagnant} is false, the loop bypasses the Teacher and returns directly to the Student (dashed edge); if true, the Teacher LLM is invoked and its guidance $\mathcal{G}$ biases the next Student call, matching Algorithm~\ref{alg:teach-to-crash}.}
    \Description{Left-to-right workflow diagram. Inputs feed the Student LLM, which outputs a JSON batch of scenarios to the Scenario Validator, which passes valid scenarios to the Simulator and system under test. The simulator produces outcomes that are appended to the Outcome History. The Outcome History feeds a Stagnation Detector. If not stagnant, the loop returns directly to the Student, bypassing the Teacher. If stagnant, the Stagnation Detector invokes the Teacher LLM, which returns guidance to the Student.}
    \label{fig:workflow}
\end{figure*}

To address the problem outlined in Section~\ref{sec:problem}, we propose \textit{Teach-to-Crash}, a closed-loop Student--Teacher LLM framework for generating critical test scenarios for ADSs, built around a stagnation-detection mechanism that monitors rolling collision-rate and time-to-collision progress and triggers Teacher intervention only once that progress plateaus (Section~\ref{sec:monitoring}). Figure~\ref{fig:workflow} illustrates the framework, and Algorithm~\ref{alg:teach-to-crash} summarizes its workflow. The framework treats the ADS under test as a black box, requiring only simulator execution and outcome logging.
It takes as input the ADS-simulator integration, a design space over NPC counts, speeds, and ego-centric positions, and $\mathrm{CR}$/$\mathrm{TTC}$ targets~\cite{Li2024TowardsGP}, then iterates generation, validation, execution, and stagnation-aware guidance.

\begin{algorithm}[t]
\caption{Student--Teacher LLM-Driven Search}
\label{alg:teach-to-crash}
\small
\LinesNumbered
\KwIn{$ADS$, $Map$; $B$: Student batch size; $I_{\max}$: max iterations}
\KwIn{$W$: rolling window; $\mathrm{CR}_{\text{target}}, \mathrm{TTC}_{\text{target}}$: intervention targets}
\KwOut{$\mathcal{H}$: outcome history (scenario archive); $\mathcal{S}_{best}$: best critical scenarios}
$\mathcal{H}\gets \emptyset$; $\mathcal{G}\gets \emptyset$; $\mathcal{S}_{best}\gets \emptyset$\;
\For{$i\gets 1$ \KwTo $I_{\max}$}{
    $\mathcal{B}\gets \textsc{StudentGenerate}(B,\mathcal{G},\theta)$\tcp*[f]{Generate scenarios}
    \ForEach{$s\in \mathcal{B}$}{
        $o\gets \textsc{ExecuteSim}(s,Map,ADS)$\tcp*[f]{Simulate scenario}
        $\mathcal{H}\gets \mathcal{H}\cup\{(s,o)\};$\tcp*[f]{record result}
        $\mathcal{S}_{best}\gets \textsc{UpdateBest}(\mathcal{S}_{best},s,o);$\tcp*[f]{Retain best}
    }
    $(\mathrm{CR}_W,\mathrm{TTC}_W)\gets \textsc{RollingMetrics} (\mathcal{H},W)$\tcp*[f]{Compute collision metrics}\\
    \eIf{$\textsc{Stagnant}(\mathrm{CR}_W,\mathrm{TTC}_W)$}{
        $\mathcal{G}\gets \textsc{TeacherGuide}(\mathcal{H},\mathrm{CR}_W,\mathrm{TTC}_W)$\tcp*[f]{Rewrite strategy}
    }{
        $\mathcal{G}\gets \emptyset$\;
    }
    \If{$\textsc{BudgetExhausted}()$}{
        \textbf{break}\;
    }
}
\Return{$\mathcal{H}, \mathcal{S}_{best}$}\;
\end{algorithm}

\begin{figure}[t]
\centering
\includegraphics[width=\columnwidth]{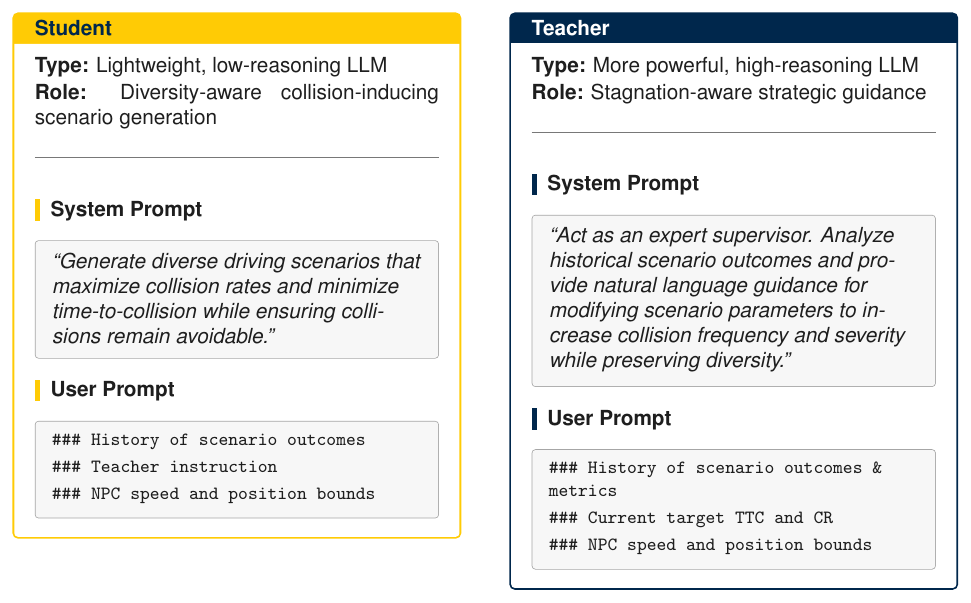}
\caption{Prompt templates for the Student and Teacher components, showing each role's type, objective, system prompt, and user-prompt structure.}
\Description{Two side-by-side prompt-template cards. The Student card lists type as lightweight, low-reasoning LLM; role as diversity-aware collision-inducing scenario generation; a system prompt instructing generation of diverse, collision-maximizing, avoidable scenarios; and a user-prompt template with placeholders for history of scenario outcomes, Teacher instruction, and NPC speed and position bounds. The Teacher card lists type as more powerful, high-reasoning LLM; role as stagnation-aware strategic guidance; a system prompt instructing the model to act as an expert supervisor and provide guidance that increases collision frequency and severity while preserving diversity; and a user-prompt template with placeholders for history of scenario outcomes and metrics, current target TTC and CR, and NPC speed and position bounds.}
\label{fig:prompt-templates}
\end{figure}

\textbf{(1) Student LLM:} a lightweight, low-reasoning-tier scenario generator. It takes the most recent $10$ records in history $\mathcal{H}$, the batch size $B$, the NPC position/speed bounds, and any Teacher guidance $\mathcal{G}$, and outputs $B$ candidate scenarios as strict JSON over the encoding of Section~\ref{sec:scenario_representation}. Its prompt encodes four objectives—maximize collision rate, minimize TTC, prefer avoidable collisions (Definition~\ref{def:avoidability}), and maintain novelty/diversity—alongside hard validity constraints, the history window, and any active guidance; the history and guidance are natural-language instructions only, not rolling online signals, and neither gates Teacher intervention (Section~\ref{sec:monitoring}). Given bounds position~$\in[1,8]$/speed~$\in[0,60]$\,mph and $B{=}10$, a representative (illustrative) output is the JSON array of Eq.~\eqref{eq:example_scenario}, where \texttt{npc0} converges from position~$5$ to $7$ and \texttt{npc1} from position~$2$ to $4$, each with an accelerate-then-decelerate speed profile.

\textbf{(2) Teacher LLM:} a more capable, high-reasoning-tier strategic search controller rather than a second generator. It takes $\mathcal{H}$ (the most recent $10$ records), the NPC bounds, the rolling metrics $\mathrm{CR}_W$/$\mathrm{TTC}_W$, and the targets $\mathrm{CR}_{\text{target}}$/$\mathrm{TTC}_{\text{target}}$, and is invoked only when \textsc{Stagnant} holds (Eq.~\eqref{eq:stagnation}). It outputs one line of guidance $\mathcal{G}$ that biases the next Student generation call—per-step position trajectories, speed profiles, NPC role assignments (aggressor vs.\ blocker), and validity reminders—warranting its stronger tier since it is called less often but must reason over recent search dynamics, e.g., directing \texttt{npc0} to converge front-center while \texttt{npc1} shifts inward to block lateral escape. Figure~\ref{fig:prompt-templates} summarizes each role's type, objective, and system/user prompt structure; full prompt templates are in the replication package~\cite{ReplicatioPackageZenodo}.

\textbf{(3) Scenario Validator:} a deterministic checking layer between the Student's outputs and the simulator. It parses the raw JSON, verifies each scenario has exactly the two expected NPC keys with $K{=}6$ $[v_k,p_k]$ pairs, clips speeds/positions to the bounds of Section~\ref{sec:scenario_representation}, and discards scenarios failing this check. The \texttt{npc0}/\texttt{npc1} distinct-position instruction is a prompt-level constraint not re-verified by the validator (Section~\ref{sec:scenario_representation}); overlap avoidance thus relies on LLM compliance.

Each iteration executes \textsc{StudentGenerate}, followed by \textsc{ExecuteSim}, which records outcomes in $\mathcal{H}$ and updates the best scenarios via \textsc{UpdateBest}, and then \textsc{RollingMetrics}. If \textsc{Stagnant} holds, \textsc{TeacherGuide} is invoked before the next iteration. The framework finally returns $\mathcal{H}$ together with the best critical scenarios discovered.

\subsection{Operating Environment}
Scenario simulation runs in CARLA (v0.9.12) using synchronous simulation and the CARLA Python API, with simulator-specific logic isolated in the interface layer to keep the architecture simulator-agnostic. The interface comprises a compiler-adapter and an execution manager. The compiler-adapter translates symbolic scenario programs into simulator API calls, spawns ego/NPC actors, attaches sensors, decodes each \texttt{position\_id} into an ego-relative waypoint target, and places each NPC at its first decoded location so it begins moving toward its first commanded $[v_k,p_k]$ pair rather than from rest. The execution manager resets the environment, runs a fixed-horizon 60\,s synchronous episode that advances the scenario index and each NPC's active pair every 10\,s, and terminates early on collision or otherwise at the horizon. Each response is parsed as a raw JSON array. Unparseable responses and scenarios that fail validation are discarded and are not added to $\mathcal{H}$. Consequently, neither LLM receives feedback about which candidates failed or the reasons for their rejection. Instead, both models observe the downstream execution outcomes of validated scenarios through the rolling history included in their prompts, including collision occurrence, the resulting TTC, and, for collision cases, the avoidability label defined in Definition~\ref{def:avoidability}.

\textbf{Collision detection and NPC--NPC interactions.} An ego-mounted sensor captures collision events; only ego--registered-NPC contacts count as scenario-relevant, and rear-side ego collisions may also be excluded depending on configuration. We estimate fault attribution (ego vs.\ NPC) heuristically from the ego's speed at contact and the relative quadrant of the colliding NPC. Because the sensor is ego-mounted, NPC--NPC contact is not separately sensed, and the validator's distinct-position instruction reduces, but does not code-enforce, overlapping targets, so continuous-time avoidance while transitioning between cells is not guaranteed.


\subsection{Scenario Representation}
\label{sec:scenario_representation}

\textit{Teach-to-Crash} represents each scenario as a simulator-executable program with two NPCs (\texttt{npc0}, \texttt{npc1}). Each NPC is a length-$K{=}6$ sequence of $[v_k,p_k]$ pairs, where $v_k$ is the target speed (mph) and $p_k$ is a discrete ego-centric \texttt{position\_id}. The Student outputs a raw JSON array (no extra text) whose elements map \texttt{npc0} and \texttt{npc1} to six such pairs, spanning the rollout's six 10\,s segments. An example appears in Equation~\ref{eq:example_scenario}.
\begingroup
\scriptsize
\begin{equation}
\label{eq:example_scenario}
\begin{aligned}
\texttt{[\{"npc0":[[32,5],[40,5],[45,6],[38,6],[30,7],[25,7]],}\\
\texttt{"npc1":[[28,2],[35,3],[42,3],[40,4],[34,4],[30,4]]\}]}
\end{aligned}
\end{equation}
\endgroup

\textbf{Validity constraints.} Before execution, the deterministic validator of Section~\ref{sec:framework_overview} validates each scenario and clips it to admissible bounds. Our case study uses an ego-centric grid with $N=9$ cells including ego: speeds clip to $[0,60]$ mph, and positions clip to integer IDs in $[1,8]$ (NPC regions), with cell~9 (ego) disallowed for NPCs. The validator also enforces exactly two NPC keys and $K{=}6$ pairs per NPC, discarding scenarios that fail this check. Both LLM prompts additionally instruct that \texttt{npc0} and \texttt{npc1} not share a \texttt{position\_id} at the same step, but this constraint is requested at generation time only and is not re-checked by the validator.

\textbf{Design rationale.} The ``low-level parameterizations'' contrasted in Section~\ref{sec:intro} are raw, continuous, simulator-native controls; ours, though still numeric, is a small, fixed-alphabet, ego-relative grammar adopted over absolute coordinates or free-form placement because it (i) removes the need to reason about map-specific coordinates or headings, since every target is relative to the ego (e.g., ``front-left''), and (ii) constrains output to a compact grammar the validator can check and repair cheaply. This relates conceptually to Frenet-frame representations, which similarly re-express agent state relative to a reference path. We choose $K{=}6$ steps to divide the rollout into six 10\,s segments, sufficient for an approach, interaction, and resolution phase; we did not sweep alternative $K$, leaving this for future work (Section~\ref{sec:discussion}).

\textbf{Relation to prior representations.} Unlike language-based representations that describe an interaction in free-form or templated text (e.g., ChatScene~\cite{ChatScene}, LeGEND~\cite{LeGEND2024}), ours constrains the LLM to directly emit points in a fixed grammar. Our grid follows the same 9-position, ego-centric convention as PAFOT~\cite{pafotpaper} and AVFuzzer~\cite{li2020av} but replaces their genetic/fuzzing search with LLM-driven generation and Teacher-guided rewriting; unlike DoFuzz~\cite{ji2025autonomous}, diversity (Section~\ref{sec:monitoring}) is prompt-enforced rather than a search operator, and unlike Feng et al.~\cite{feng2021intelligent}'s dense, adaptive importance-sampling representation, our grid trades that adaptivity for a smaller, LLM-tractable space.

\textbf{Scope.} The present evaluation exercises this representation on a single class of highway-lane interactions between an ego vehicle and two NPCs (Section~\ref{sec:exp-setup}). We do not evaluate intersections, turns, or general urban traffic-rule interactions, nor do we claim generalization to those classes.

\subsection{Search Objectives}
\label{sec:monitoring}

To assess whether generated scenarios achieve the objectives of Section~\ref{sec:problem}, we monitor two quantities online over a rolling window of the most recent $W$ scenarios: collision rate, which captures how often the search finds failure-inducing scenarios, and rolling TTC, the primary online criticality proxy. We measure diversity offline instead, over the full archive, and also archive $\mathrm{Crit}(\theta)$ (Definition~\ref{def:criticality}) for offline ranking. We use the term \emph{prompt-level diversity} for the natural-language instructions given to both LLMs (e.g., ``avoid near-duplicate scenarios,'' ``keep at least one-third of the batch clearly novel'')---a soft, unenforced signal distinct from \emph{parameter distance}, the measure used to compute diversity offline: the Jaccard-based pairwise distance over the position/speed-bin fingerprints $F(\theta)$ of Definition~\ref{def:diversity} (Section~\ref{sec:eval}).

Let $\mathcal{S}_W = \{s_1,\dots,s_W\}$ denote the most recent $W$ executed scenarios; the rolling collision rate and TTC are
{\small
\begin{equation}
\label{eq:rolling_metrics}
\text{CR}_W \;=\; \frac{1}{W}\sum_{i=1}^{W}\mathbb{I}[\text{collision}_i],
\qquad
\text{TTC}_W \;=\; \frac{1}{W}\sum_{i=1}^{W}\text{TTC}_i.
\end{equation}
}
Here, $\text{collision}_i \in \{0,1\}$ indicates whether $s_i$ collided, and $\text{TTC}_i$ is the minimum observed time-to-collision, or its estimated proxy when no collision occurs; the search thus aims to increase $\text{CR}_W$ while decreasing $\text{TTC}_W$. Diversity, being $O(n^2)$ to compute (Definition~\ref{def:diversity}), is reserved for post-hoc analysis over the full archive $G$ rather than computed online; the most critical scenarios recorded each iteration likewise support reproducibility, while rolling collision/TTC remain the online signal triggering Teacher intervention.


Let $\text{CR}_W^{(t)}$ and $\text{TTC}_W^{(t)}$ denote the rolling metrics after iteration $t$, once at least $W$ scenarios have been executed. Rather than comparing these to the immediately preceding window, the implementation compares them against the most recent \emph{improving} checkpoint, $\text{CR}_W^{(\text{ref})}$ and $\text{TTC}_W^{(\text{ref})}$—recorded the last time the search was judged to be progressing—so stagnation reflects a plateau since the last confirmed improvement, not a single-step decline. Progress on each objective is judged by reaching its target, or by improving on the reference by at least a small tolerance $\varepsilon$:
{\small
\begin{align}
\label{eq:stagnation}
\text{CollImproving}^{(t)} &\equiv \bigl(\text{CR}_W^{(t)} \ge \text{CR}_{\text{target}}\bigr) \lor \bigl(\text{CR}_W^{(t)} - \text{CR}_W^{(\text{ref})} \ge \varepsilon_{\text{CR}}\bigr),\notag\\
\text{TTCImproving}^{(t)} &\equiv \bigl(\text{TTC}_W^{(t)} \le \text{TTC}_{\text{target}}\bigr) \lor \bigl(\text{TTC}_W^{(\text{ref})} - \text{TTC}_W^{(t)} \ge \varepsilon_{\text{TTC}}\bigr),\\
\textsc{Stagnant}^{(t)} &\equiv \lnot\bigl(\text{CollImproving}^{(t)} \land \text{TTCImproving}^{(t)}\bigr),\notag
\end{align}
}
In our configuration, $\varepsilon_{\text{CR}}=0.02$ and $\varepsilon_{\text{TTC}}=1.0$\,s: when both objectives are improving, the reference checkpoint updates to the current window and no Teacher call is issued; otherwise $\textsc{Stagnant}^{(t)}$ holds, the Teacher returns guidance $\mathcal{G}$ biasing the next $\textsc{StudentGenerate}$ call toward revised positions, timing, and speed evolution, and the reference stays unchanged until progress resumes. Algorithm~\ref{alg:teach-to-crash} presents this as the simplified predicate $\textsc{Stagnant}(\mathrm{CR}_W,\mathrm{TTC}_W)$; Eq.~\eqref{eq:stagnation} is the exact criterion evaluated.

\section{Evaluation Setup}
\label{sec:exp-setup}
We demonstrate \textit{Teach-to-Crash} in CARLA~\cite{Dosovitskiy2017CARLA}, a high-fidelity simulator widely used for developing and testing ADS. We use the Town06 map, which combines urban and highway segments (highway junctions, long straightaways, intersections, merging/diverging lanes, and Michigan Left turns) to enable rich high-speed interactions. The CARLA server runs in synchronous mode with a fixed $0.05$s simulation time step for deterministic physics and a $60$s simulation horizon, balancing interaction complexity against computational efficiency.

Each scenario contains three vehicles. The ego vehicle is sampled from the CARLA blueprint library and controlled by CARLA's built-in autopilot/Traffic Manager, the primary evaluated controller. The two adversarial agents (NPC0, NPC1) are Nissan Micra models whose initial positions and velocities are fully determined by the Student model via quadrant-based spawning. For our main effectiveness, timing, and diversity evaluation (Section~\ref{sec:eval}), we use the autopilot/Traffic Manager for fair comparison against the baselines. For usefulness (Section~\ref{subsec:rq4}), we additionally evaluate \textit{Teach-to-Crash} scenarios using the more advanced, learning-based TransFuser controller~\cite{Chitta2023PAMI}.

We define two experimental setups that share the same map, NPC configuration, and operational design domain but differ in the ego controller's speed policy, enabling comparison under different levels of ego aggressiveness. \emph{Setup~A} uses the Traffic Manager at the posted speed limit $60$\,mph. \emph{Setup~B} reduces the ego's target speed by $20\%$, to approximately $48$\,mph. Both setups set the global following distance to $5.0$\,m.

The shared operational design domain (ODD) assumes sunny daytime weather and standard traffic and road components (traffic lights, lane markers, pedestrian crossings), with four-to-six-lane road configurations and a $60$\,mph speed limit for both the ego and NPCs. We fix the number of adversarial NPCs at two, sufficient for meaningful interactions without overcrowding the ego's vicinity.

We instantiate the Student with \textit{gpt-5-nano} and the Teacher with \textit{gpt-5-mini}, balancing reasoning capability against cost and latency. We evaluate this configuration over $10$ independent runs, each consisting of $I=45$ iterations.\footnote{Experiments were conducted on Windows 11 using an Intel Core i9 processor with 64 GB of RAM and an NVIDIA RTX 3070 GPU.} Each iteration generates $10$ candidate scenarios, and Teacher guidance uses rolling statistics over the most recent $10$ records. The full implementation, evaluation data, and results are available in the replication package~\cite{ReplicatioPackageZenodo}.

\section{Evaluation}
\label{sec:eval}
This section evaluates \textit{Teach-to-Crash} in the CARLA case study described in Section~\ref{sec:exp-setup}: collision-finding effectiveness against baseline methods, diversity of generated test scenarios, and usefulness for downstream verification. The replication package is publicly available on GitHub~\cite{ReplicatioPackageZenodo}.

\subsection{Baseline Methods}
\label{subsec:baselines}

We compare \textit{Teach-to-Crash} against two baselines spanning distinct paradigms: optimization-based search, named PAFOT, and knowledge-driven LLM generation, named ChatScene. \textbf{PAFOT~\cite{pafotpaper}} is our conventional non-LLM baseline, which is a genetic algorithm that searches a fixed ego-centric grid, evolving scenarios toward lower estimated TTC and minimum distance. \textbf{ChatScene~\cite{ChatScene}} is our leading open-loop LLM baseline, using retrieval-augmented generation to transform natural-language descriptions into executable Scenic scripts. Across all three methods, we use the same CARLA map, ego controller, NPC types, synchronous simulator configuration, 60\,s horizon, and collision logging. For \textit{Teach-to-Crash}, the Teacher targets $\mathrm{CR}_{\text{target}} = 90\%$ and $\mathrm{TTC}_{\text{target}} = 12$\,s, declaring stagnation when the rolling window shows no improvement toward either.


\subsection{Research Questions}
\label{subsec:rqs}
We structure the evaluation around the following research questions:

\noindent\textbf{RQ1 (Effectiveness):} How effectively does \textit{Teach-to-Crash} generate collision-inducing scenarios compared to baseline methods? 

\noindent\textbf{RQ2 (Collision Timing/Criticality):} When collisions occur, how quickly, i.e., how early within the fixed scenario horizon, does \textit{Teach-to-Crash} induce them compared to baseline methods?

\noindent\textbf{RQ3 (Diversity):} Does \textit{Teach-to-Crash} generate a more diverse set of collision-inducing scenarios than the baseline methods?

\noindent\textbf{RQ4 (Avoidability-Based Usefulness Proxy):} Does \textit{Teach-to-Crash} produce more collisions labeled avoidable, under the avoidability heuristic of Definition~\ref{def:avoidability}, than the baseline methods?
\newline
\newline
RQ1 measures effectiveness via Collision Hit Rate (CHR) and Collision Discovery Rate (CDR) over 10 runs. RQ2 measures collision timing via Time-to-Collision (TTC), the elapsed simulation time from rollout start to collision: a lower TTC indicates an earlier, more critical failure within the fixed 60\,s horizon, not faster or cheaper testing, since it excludes wall-clock time, LLM latency, and scenario validation cost (Section~\ref{sec:discussion}). CDR normalizes by executed simulator time and is our closest proxy to throughput. RQ3 quantifies diversity via Jaccard novelty, failure clusters, and unique-failure ratio (Definition~\ref{def:diversity}); RQ4 measures usefulness via the avoidability-based proxy (Definition~\ref{def:avoidability}): avoidable collision rate, collisions per run, and per 100 scenarios. All experiments are repeated 10 times, with significance assessed via Mann-Whitney U tests with Benjamini-Hochberg correction~\cite{mann1947test}.

\subsection{RQ1: Collision-Finding Effectiveness}
\label{subsec:rq1}

We evaluate collision-finding effectiveness using two standard metrics. The
\emph{Collision Hit Rate} (CHR), defined in Eq.~\eqref{eq:chr}, is the percentage
of executed scenarios resulting in a detected collision:

\begin{equation}
\label{eq:chr}
\mathrm{CHR} =
\frac{\#\mathrm{Collisions}}
{\#\mathrm{Total}}
\times 100.
\end{equation}

\begin{figure*}[t]
\centering
\includegraphics[width=\textwidth]{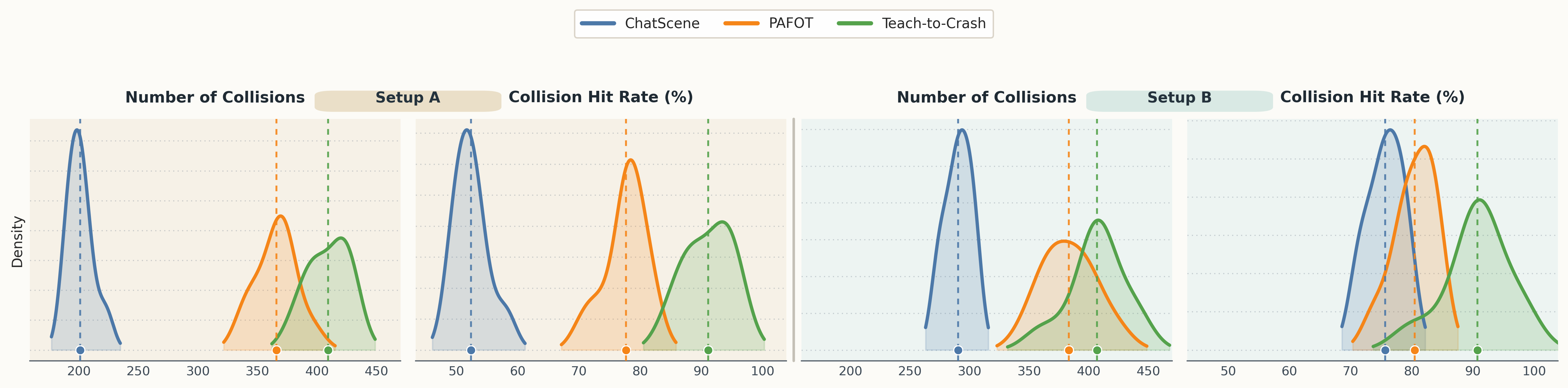}
\caption{Kernel density estimates (KDE) of per-method collision counts and collision hit rate across Setups A and B, each fit to $10$ run-level observations. Dashed lines mark method means.}

\Description{A four-panel figure in one row with smooth density plots for number of collisions and collision hit rate in Setup A and Setup B. ChatScene, PAFOT, and Teach-to-Crash are shown with distinct colored curves and dashed lines marking their means.}
\label{fig:rq1-metrics-smooth}
\end{figure*}

The \emph{Collision Discovery Rate} (CDR), defined in Eq.~\eqref{eq:cdr},
measures the number of detected collisions per hour of executed simulator time:

\begin{equation}
\label{eq:cdr}
\mathrm{CDR} =
60 \times
\frac{\#\mathrm{Collisions}}
{\sum_{\theta\in G} T_{\mathrm{exec}}(\theta)},
\end{equation}

where $T_{\mathrm{exec}}(\theta)$ denotes the executed simulator time of
scenario $\theta$, in minutes. Figure~\ref{fig:rq1-metrics-smooth} shows the
distribution of results, and the full per-run table is available in the
replication package~\cite{ReplicatioPackageZenodo}.

\textit{Teach-to-Crash} achieves the highest CHR in both setups: $91.10 \pm 3.84\%$ (Setup~A) and $90.48 \pm 3.20\%$ (Setup~B), versus $77.67$/$80.50\%$ for PAFOT and $52.40$/$75.65\%$ for ChatScene. On CDR, PAFOT attains higher means ($161.81$ and $197.06$) than \textit{Teach-to-Crash} ($127.68$ and $144.74$), but with substantially larger variance ($\pm 69.12$ vs.\ $\pm 14.99$ in Setup~A), driven by a few runs exploiting a narrow fast-to-collide region. A plausible explanation is that PAFOT's genetic search operates directly over the same 9-position ego-centric grid (Section~\ref{sec:related}) and, in a subset of runs, converges on generations that place an NPC immediately adjacent to the ego, producing a burst of near-instantaneous collisions that inflates that run's CDR; because this depends on which grid regions a given run's population happens to explore, the effect appears in only some runs rather than uniformly, which is consistent with a high-mean, high-variance pattern instead of a consistently higher rate. \textit{Teach-to-Crash} shows a higher per-scenario collision likelihood across the evaluated runs.

\begin{table}[t]
\centering
\scriptsize
\setlength{\tabcolsep}{4.2pt}
\renewcommand{\arraystretch}{0.98}
\caption{Teach-to-Crash ablation across Setup A and B. Student-only: scenarios without Teacher rewriting; Teacher-guided: scenarios after Teacher intervention.}
\label{tab:rq1-teacher-ablation}
\resizebox{0.48\textwidth}{!}{%
\begin{tabular}{clcccc}
\toprule
\textbf{Setup} & \textbf{Variant} & \textbf{Scen.} & \textbf{Coll.} & \textbf{CHR (\%)} & \textbf{CDR} \\
\midrule
\multirow{3}{*}{\textbf{A}}
& \textit{Teach-to-Crash} & 4493 & 4093 & $91.10 \pm 3.84$ & $127.68 \pm 14.99$ \\
& Student only & 1000 & 929 & $92.58 \pm 3.87$ & $29.37 \pm 9.62$ \\
& Teacher-guided & 3493 & 3164 & $90.78 \pm 4.38$ & $98.31 \pm 6.89$ \\
\midrule
\multirow{3}{*}{\textbf{B}}
& \textit{Teach-to-Crash} & 4498 & 4070 & $90.48 \pm 3.20$ & $144.74 \pm 9.73$ \\
& Student only & 470 & 432 & $91.83 \pm 4.08$ & $15.44 \pm 3.92$ \\
& Teacher-guided & 4028 & 3638 & $90.34 \pm 3.23$ & $129.30 \pm 7.49$ \\
\bottomrule
\end{tabular}%
}
\end{table}

Table~\ref{tab:rq1-teacher-ablation} further ablates \textit{Teach-to-Crash} into \emph{Student-only} proposals (no Teacher intervention) and \emph{Teacher-guided} proposals. In Setup A, the Student alone reaches a high CHR on the subset it executes ($92.58 \pm 3.87\%$), but yields only $929$ collisions overall ($29.37 \pm 9.62$/hr), while Teacher-guided scenarios account for $3164$ of $4093$ total collisions ($77.3\%$, $98.31 \pm 6.89$/hr); Setup B shows the same pattern, with Teacher-guided scenarios accounting for $3638$ of $4070$ total collisions. This quantifies the Teacher's \emph{aggregate} contribution within the reported Student-only and Teacher-guided scenario partition, not intervention timing or diversity evolution, left to future work (Section~\ref{sec:discussion}).

\begin{tcolorbox}[breakable,boxsep=0pt,left=3pt,right=3pt,colback=white]
\textbf{RQ1:} \textit{Teach-to-Crash} achieves the highest Collision Hit Rate averaged across both setups ($90.79\%$), significantly outperforming PAFOT ($79.09\%$) and ChatScene ($64.03\%$). While PAFOT exhibits a higher mean Collision Discovery Rate ($179.44$ col/hr vs.\ $136.21$), \textit{Teach-to-Crash} exhibits lower CDR variance in Setup A. Ablation confirms Teacher-guided scenarios account for over $77\%$ of all collisions discovered within the reported Student-only and Teacher-guided scenario partition, an aggregate association that does not isolate intervention timing or causal contribution.
\end{tcolorbox}

\subsection{RQ2: Collision Timing and Criticality}
\label{subsec:rq2}
\begin{figure}[t]
\centering
\includegraphics[width=\linewidth]{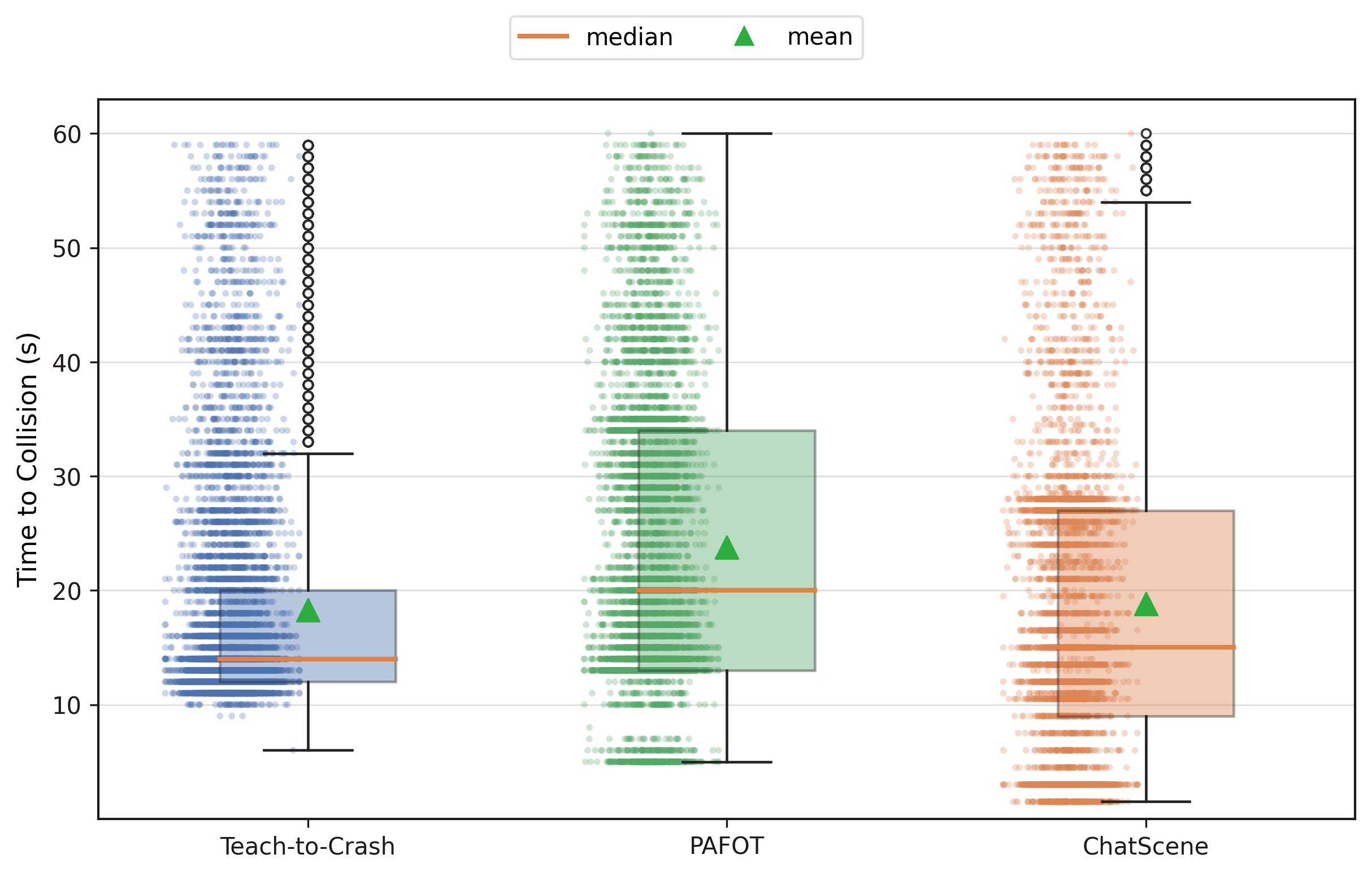}
\caption{Times-to-collision for each method.}
\Description{Boxplots comparing time-to-collision distributions for ChatScene, PAFOT, and Teach-to-Crash. Teach-to-Crash has the lowest median and mean time-to-collision, followed by PAFOT, with ChatScene highest.}
\label{fig:ttc-boxplot}
\end{figure}

We examine Time-to-Collision (TTC) for scenarios ending in collision, summarized as boxplots in Figure~\ref{fig:ttc-boxplot}; lower TTC indicates an earlier collision within the fixed horizon, a timing/criticality signal rather than a testing-speed measure. \textit{Teach-to-Crash} yields the shortest TTC overall: run-level mean $17.41$\,s, versus $22.42$\,s (PAFOT) and $30.32$\,s (ChatScene). The same advantage over PAFOT holds in Setup B: $19.20 \pm 0.77$\,s versus $25.00 \pm 1.52$\,s, about $23.2\%$ lower.

We treat shorter TTC as a criticality signal, not evidence of end-to-end efficiency: run length is controlled by the fixed iteration budget $I_{\max}$ rather than early termination, so we do not measure wall-clock savings. \textit{Teach-to-Crash} scenarios often exhibit tighter initial NPC--ego spacing within the first few steps, whereas ChatScene often places adversarial agents at positions requiring long approach phases, inflating TTC even when a collision eventually occurs.

\label{subsec:statistical-significance}
We apply two-sided Mann--Whitney U tests~\cite{mann1947test} to per-run CHR, CDR, and TTC, reporting bootstrap 95\% CIs on run-level means with rank-biserial effect size $r_{rb}$. Setup~A CHR: \textit{Teach-to-Crash} $91.10$ [$88.81, 93.27$] vs.\ $77.67$ [$75.62, 79.42$] (PAFOT) and $52.40$ [$50.96, 54.09$] (ChatScene). TTC: $17.41$\,s [$16.67, 18.09$] vs.\ $22.42$\,s [$21.37, 23.38$] and $30.32$\,s [$29.95, 30.65$]. CDR: $127.70$ [$119.29, 136.73$], $161.81$ [$134.19, 207.43$], $42.49$ [$40.88, 44.33$]. \textit{Teach-to-Crash} significantly outperforms both baselines on CHR and TTC ($p<0.001$, $|r_{rb}|=1.00$) and ChatScene on CDR ($p<0.001$, $r_{rb}=1.00$), though the CDR difference against PAFOT is not significant ($p=0.063$, $r_{rb}=-0.50$). All Setup~B pairwise differences are significant ($p<0.001$); full comparisons are in the replication package~\cite{ReplicatioPackageZenodo}.


\begin{tcolorbox}[breakable,boxsep=0pt,left=3pt,right=3pt,colback=white]
\textbf{RQ2:}
\textit{Teach-to-Crash} achieves the shortest Time-to-Collision averaged across both setups, with a mean of $18.31$\,s versus $23.71$\,s for PAFOT and $30.32$\,s for ChatScene, reflecting an earlier collision within the rollout horizon rather than faster or cheaper testing.
\end{tcolorbox}

\subsection{RQ3: Diversity of Collision Scenarios}
\label{subsec:rq3}

We compare diversity across 10 runs using three metrics built on the scenario fingerprints $F(\theta)$ from Definition~\ref{def:diversity}: \emph{Jaccard novelty} (how different each new scenario is from prior ones in the same run); \emph{failure clusters} (the number of \emph{distinct} fingerprints among collision-inducing scenarios in a run); and \emph{unique-failure ratio} (cluster count normalized by scenarios executed). These summaries do not establish whether failure clusters are shared or unique across methods, an analysis left for future work (Section~\ref{sec:discussion}).

\textit{Teach-to-Crash} leads on all three measures in both setups. In Setup~A it reaches a Jaccard novelty of $0.539 \pm 0.056$, $404.7 \pm 15.4$ failure clusters, and a unique-failure ratio of $0.901 \pm 0.035$, versus ChatScene ($0.522$, $199.2$, $0.519$) and PAFOT ($0.336$, $236.3$, $0.501$). The same ranking holds in Setup~B ($0.554$, $404.0$, $0.898$ vs.\ $0.522$/$289.1$/$0.753$ for ChatScene and $0.339$/$255.2$/$0.536$ for PAFOT). ChatScene maintains high proposal novelty but collapses onto fewer collision outcomes, while PAFOT repeatedly rediscovers nearby failures.

\begin{figure}[t]
    \centering
    \includegraphics[width=1\columnwidth]{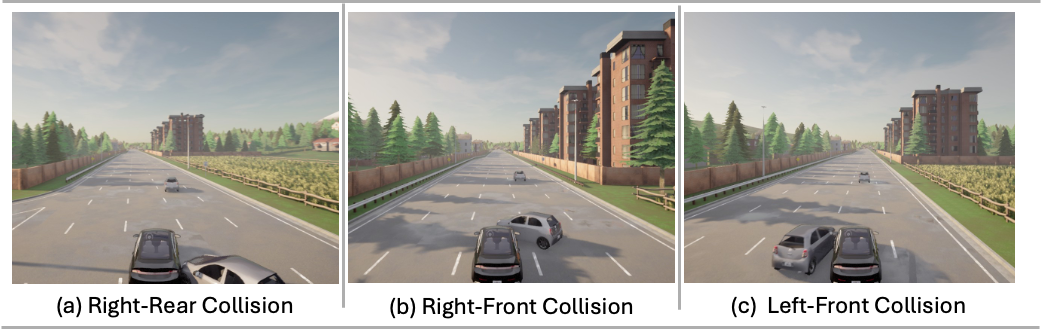}
    \caption{Three collision scenarios discovered by \textit{Teach-to-Crash}, each exhibiting a distinct ego--NPC interaction pattern: (a) right-rear collision, (b) right-front collision, and (c) left-front collision.}
    \Description{Three side-by-side screenshots from the CARLA simulator showing different collision scenarios on a multi-lane highway in Town06: (a) a right-rear collision, (b) a right-front collision, and (c) a left-front collision between the ego vehicle and an NPC.}
    \label{fig:collision-examples}
\end{figure}

Figure~\ref{fig:collision-examples} shows three representative \textit{Teach-to-Crash} collision scenarios (right-rear, right-front, left-front), illustrating structurally distinct interaction types rather than parameter variations of a single pattern; these are illustrative archive examples, not evidence about overlap with baseline patterns.

\smallskip\noindent\textit{Discussion.}
The CHR/CDR ablation (Table~\ref{tab:rq1-teacher-ablation}) shows Teacher-guided scenarios account for most collisions within the reported Student-only and Teacher-guided scenario partition; this aggregate association neither shows that intervention causes greater diversity nor measures diversity directly, so we leave a dedicated diversity ablation to future work. PAFOT's fixed genetic operators lack awareness of unexplored scenario-space regions, while ChatScene's linguistic variability often compiles to similar executable behaviors, yielding moderate novelty but low failure-cluster diversity.

\begin{tcolorbox}[breakable,boxsep=0pt,left=3pt,right=3pt,colback=white]
\textbf{RQ3:} \textit{Teach-to-Crash} achieves the highest average Jaccard novelty ($0.547$) and discovers the most failure clusters ($404.4$ per run) averaged across both setups, showing broader collision-pattern coverage than both baselines. Its average unique-failure ratio of $0.900$ shows over $90\%$ of executed scenarios introduce novel collision signatures, substantially less redundant than ChatScene ($0.636$) and PAFOT ($0.519$).
\end{tcolorbox}

\subsection{RQ4: Avoidability-Based Usefulness Proxy}
\label{subsec:rq4}

We evaluate avoidability as a proxy for whether discovered failures may be useful for downstream failure analysis in ADS, not as a direct measurement of debugging usefulness. A collision is labeled \emph{avoidable} if, at any sampled moment in a pre-collision window, either (i) there is at least $5$\,m of lateral space, or (ii) the space ahead exceeds $d_{\text{stop}} = v^2/(2a) + 1.5$ with $a=6.5$\,m/s$^2$. We report \emph{avoidable collision rate} (\%), \emph{per run}, and \emph{per 100 executed scenarios}, and also evaluate TransFuser TraverseV4~\cite{Chitta2023PAMI}, a learning-based controller, alongside CARLA's built-in autopilot/Traffic Manager, as an exploratory check under a different controller.

\begin{table*}[t]
\centering
\scriptsize
\setlength{\tabcolsep}{11pt}
\renewcommand{\arraystretch}{1}
\caption{Usefulness results across Setup A and B: avoidable-collision rate, avoidable collisions per run and per 100 executed scenarios.}
\label{tab:rq4-usefulness}
\resizebox{\textwidth}{!}{%
\begin{tabular}{clccc}
\toprule
\textbf{Setup} & \textbf{Method} & \textbf{Avoidable Collision Rate (\%)} & \textbf{Avoidable Coll./Run} & \textbf{Avoidable Coll./100 Scen.} \\
\midrule
\multirow{4}{*}{\textbf{A}}
& ChatScene & 23.56 $\pm$ 3.51 & 47.5 $\pm$ 8.0 & 12.37 $\pm$ 2.07 \\
& PAFOT & 30.91 $\pm$ 5.91 & 113.4 $\pm$ 24.2 & 24.01 $\pm$ 4.80 \\
& \textit{Teach-to-Crash} & 39.80 $\pm$ 18.22 & 164.8 $\pm$ 81.4 & 36.69 $\pm$ 18.16 \\
& \textit{Teach-to-Crash} (TraverseV4) & 84.06 $\pm$ 6.55 & 192.0 $\pm$ 87.7 & 61.83 $\pm$ 15.24 \\
\midrule
\multirow{3}{*}{\textbf{B}}
& ChatScene & 78.65 $\pm$ 1.74 & 228.4 $\pm$ 7.7 & 59.48 $\pm$ 2.01 \\
& PAFOT & 30.80 $\pm$ 6.04 & 117.5 $\pm$ 20.8 & 24.67 $\pm$ 4.35 \\
& \textit{Teach-to-Crash} & 80.27 $\pm$ 8.34 & 326.3 $\pm$ 32.4 & 72.54 $\pm$ 7.18 \\
\bottomrule
\end{tabular}%
}
\vspace{-6pt}
\end{table*}

We evaluate TraverseV4 only under Setup~A, as an exploratory check under a different, learning-based controller; Setup~B reflects only the CARLA autopilot/Traffic Manager used elsewhere (Section~\ref{sec:exp-setup}).

Table~\ref{tab:rq4-usefulness} shows \textit{Teach-to-Crash} leads on all three usefulness measures in both setups. In Setup~A it attains $39.80 \pm 18.22\%$ avoidable collision rate, $164.8 \pm 81.4$ per run, and $36.69 \pm 18.16$ per 100 scenarios ($52.8\%$ over PAFOT, $2.97\times$ over ChatScene). Under TraverseV4 (Setup~A), the rate rises to $84.06 \pm 6.55\%$ ($192.0 \pm 87.7$/run, $61.83 \pm 15.24$/100 scenarios); we read this as evidence the signal persists under a second, learning-based controller, not proof of controller-general usefulness. In Setup~B, \textit{Teach-to-Crash} again leads ($80.27 \pm 8.34\%$, $326.3 \pm 32.4$/run, $72.54 \pm 7.18$/100 scenarios) against ChatScene ($78.65\%$, $228.4$, $59.48$) and PAFOT ($30.80\%$, $117.5$, $24.67$). We interpret the Setup~A result conservatively given its higher variance.

\begin{tcolorbox}[breakable,boxsep=0pt,left=3pt,right=3pt,colback=white]
\textbf{RQ4:} \textit{Teach-to-Crash} yields the highest avoidable collision rate under the avoidability-based usefulness proxy (Definition~\ref{def:avoidability}) in all evaluated configurations: $39.80\%$ with the CARLA Traffic Manager on Setup~A, $84.06\%$ with TraverseV4 on Setup~A, and $80.27\%$ on Setup~B. It outperforms both baselines under the Traffic Manager, the only controller evaluated for all three methods; TraverseV4 was evaluated only for \textit{Teach-to-Crash}, as an exploratory check under a different controller, so no direct baseline comparison exists there. The failures it discovers are both more numerous and more frequently labeled avoidable under this proxy.
\end{tcolorbox}

\section{Threats to Validity}
\label{sec:discussion}
\textbf{Internal Validity.}
Adapting ChatScene to our CARLA setup may introduce implementation bias that favors one technique over another. To reduce this risk, we evaluate all methods in the same CARLA Town06 environment using identical controllers, NPC types, and simulator configuration. Our conclusions regarding collision-finding effectiveness and scenario diversity are based on CHR, TTC, and fingerprint novelty, which are independent of LLM inference overhead. Although LLM-based generation is inherently stochastic, repeated runs and fixed prompt templates reduce, but do not eliminate, this source of variability.

\textbf{External Validity.}
Our case study is limited to a single map (Town06), a fixed ego-relative grid ($N{=}9$, $K{=}6$), two NPCs, and highway lane-following interactions; we do not evaluate intersections or urban traffic-rule scenarios (Section~\ref{sec:scenario_representation}). We evaluate \textit{Teach-to-Crash} under CARLA's autopilot/Traffic Manager and TransFuser-based TraverseV4, but not Apollo or Autoware, so controller generality claims are bounded to these two. Fixing two NPCs keeps the search space tractable but misses denser traffic; we also did not sweep $K$ or $N$.

\textbf{Construct Validity.}
We assess diversity via Jaccard novelty and failure clusters and usefulness via the avoidable-collision-rate proxy of Definition~\ref{def:avoidability}, a simplified kinematic proxy, not a substitute for reachability analysis or a demonstrated debugging benefit; our per-run results (Section~\ref{subsec:rq3}) do not establish whether methods discover overlapping or disjoint failure clusters.

\textbf{Scope and Baseline Coverage.}
Several comparisons fall outside our scope. First, \textit{Teach-to-Crash} is compared only against its Student-only ablation, PAFOT, and ChatScene, not a same-model or single stronger-LLM closed loop, so we claim no dual-LLM superiority over untested alternatives. Second, our baselines omit a diversity-aware technique such as BehAVExplor~\cite{cheng2023behavexplor} or DoFuzz~\cite{ji2025autonomous}. Third, we report no wall-clock time, LLM latency, or token cost, so we make no efficiency claims.

\section{Conclusion and Future Work}
\label{sec:conclusion}
We presented \textit{Teach-to-Crash}, a closed-loop dual-LLM architecture that uses execution feedback to steer collision-inducing ADS test scenario generation while preserving validity and diversity, achieving in our CARLA case study (a single map, highway lane-following, two NPCs) the highest collision rates, shortest mean Time-to-Collision, most diverse failure archive, and highest avoidability based usefulness proxy among compared methods on both controllers. Ablation links Teacher-guided scenarios to most discovered collisions, without isolating intervention timing or model capability (Section~\ref{sec:discussion}). These results support dual-LLM reasoning as adaptive search control, trading CHR for CDR relative to PAFOT (Section~\ref{subsec:rq1}) while treating TTC as a criticality signal distinct from unmeasured end-to-end efficiency (Section~\ref{subsec:rq2}). Future work includes sensitivity analysis for $N$, $K$, and $w_1,\ldots,w_4$; richer severity/diversity objectives; broader maps and ADS stacks; same-model, single-LLM, and diversity-aware baselines; token-cost measurement; and reachability-based avoidability analysis.

\noindent\textbf{Data Availability Statement.} All data, code, and artifacts supporting this paper's results are available in a public repository~\cite{ReplicatioPackageZenodo}.

\begin{acks}
This material is based upon work supported by the National Science Foundation under Grant No. 2347294. Any opinions, findings, and conclusions or recommendations expressed in this material are those of the author(s) and do not necessarily reflect the views of the National Science Foundation.
\end{acks}

\bibliographystyle{ACM-Reference-Format}
\bibliography{bibliography}

\end{document}